# Impact of the $La_2NiO_{4+\delta}$ oxygen content on the synaptic properties of the $TiN/La_2NiO_{4+\delta}/Pt$ memristive devices

*Aleksandra Koroleva[1,2], Thoai-Khanh Khuu[1,3], César Magén[4], Hervé Roussel[1], Carmen Jiménez[1], Céline Ternon[1], Elena-Ioana Vatajelu[2] and Mónica Burriel[1,*]*

[1] *Université Grenoble Alpes, CNRS, Grenoble INP, LMGP, Grenoble, France*

[2] *Université Grenoble Alpes, CNRS, Grenoble INP, TIMA, Grenoble, France*

[3] *Université Grenoble Alpes, CNRS, CEA/LETI Minatec, LTM, Grenoble, France*

[4] *Instituto de Nanociencia y Materiales de Aragón (INMA), CSIC-Universidad de Zaragoza, 50009 Zaragoza, Spain*

*\*Email: monica.burriel@grenoble-inp.fr*

**Keywords**

artificial synapse, resistive switching, $La_2NiO_{4+\delta}$, post-deposition annealing, interstitial oxygen, spiking neural network

**Abstract**

The rapid development of brain-inspired computing requires new artificial components and architectures for its hardware implementation. In this regard, memristive devices emerged as potential candidates for artificial synapses because of their ability to emulate the plasticity of the biological synapses. In this work, the synaptic behavior of the $TiN/La_2NiO_{4+\delta}/Pt$ memristive devices based on thermally annealed $La_2NiO_{4+\delta}$ films is thoroughly investigated. Using electron energy loss spectroscopy, we show that post-deposition annealing using inert (Ar) or oxidizing ($O_2$) atmospheres affects the interstitial oxygen content ($\delta$) in the $La_2NiO_{4+\delta}$ films. Electrical characterization shows that both devices exhibit long-term potentiation/depression and spike-timing-dependent plasticity (STDP). At the same time, the Ar annealed $TiN/La_2NiO_{4+\delta}/Pt$ device demonstrates filamentary-like behavior, fast switching and low energy consumption.

On the other hand, the $O_2$ annealed TiN/$La_2NiO_{4+\delta}$/Pt devices are forming-free, exhibiting interfacial-like resistive switching with slower kinetics. Finally, the simulation tools show that spiking neural network architectures with weight updates based on the experimental data achieve high inference accuracy in the digit recognition task, which proves the potential of TiN/$La_2NiO_{4+\delta}$/Pt devices for artificial synapse applications.

## 1. Introduction

Inspired by the outstanding efficiency of the human brain, hardware-implemented neuromorphic computing has drawn particular attention due to its low power consumption and high integration density [1] as a potential solution to overcome the data processing limitations set by the von Neumann bottleneck. Digital and analog neural networks are actively being developed and implemented for tasks that require learning and analysis, for example, image and speech recognition. The main advantage of a neural network (NN) is the ability to learn specific patterns and adapt the computational process. In the NN, the input neurons are connected with output neurons by synapses, with the ability to adapt their weights in response to the activity of the neurons. Among the potential neuromorphic architectures, the spiking neural network (SNN) is one of the closest to biological NNs, using spike trains to transmit information from input to output neurons, which allows spatial and temporal coding of the input signal [2]. Consequently, the synaptic weight update in SNNs can be realized through learning rules such as spike-timing dependent plasticity (STDP). In the STDP model, if the time difference $\Delta t$ between the post-synaptic spike and pre-synaptic spike is positive ($\Delta t > 0$), the weight of the synapse increases (potentiation). On the contrary, if $\Delta t < 0$, the synaptic weight will be decreased (depression) [3].

The STDP training rule uses a synaptic weight update usually represented by an exponential function [4]. It is relatively easy to implement in software, however, the hardware implementations of neuromorphic architectures require either complex arithmetic units or new-generation devices to mimic the function of biological synapses and successfully achieve bio-inspired training. Among others, memristive devices, or memristors, are considered promising candidates for the emulation of synaptic connections between the neurons since they exhibit synaptic plasticity and long-term potentiation and depression [5,6]. Typically, a memristor is based on metal-oxide-metal structures with the ability to change resistance from a high resistance state (HRS) to a low resistance state (LRS) with the application of external bias. Depending on the materials used in the device fabrication, resistive switching (RS) can happen due to the formation and rupture of the conductive filament (CF) consisting of the oxygen vacancies ($V_O^{\cdot\cdot}$) in the non-stoichiometric oxide layer [7,8], redox process at the metal/insulator interface [9], alteration in Schottky barrier width/height [10], trapping/detrapping of electrons at the defect sites [11], etc., which leads to a wide variety of the device properties [12]. It was shown that memristors can exhibit high endurance (up to ~$10^{12}$ switching cycles) [13], data retention of 10 years at 160 °C [14], and can be scaled down to 10 nm [15]. In addition, particular devices also demonstrate the ability to gradually and reversibly change the resistance between multiple levels with the application of short voltage pulses, which allows the use of memristors as artificial synapses in hardware neural networks [16]. Notably, memristive devices based on perovskite-based oxides such as $LaMnO_{3+\delta}$ [17–19], $(La,Sr)MnO_{3-\delta}$ [20,21], $Pr_{0.7}Ca_{0.3}MnO_3$ (PCMO) [22–25], or $La_2NiO_{4+\delta}$ [26–30] drew attention due to their promising analog properties, which makes them attractive candidates for synaptic applications.

Particularly, the memristive behavior of the p-type semiconductor $La_2NiO_{4+\delta}$ (L2NO4) was recently thoroughly investigated [26–30]. L2NO4 is the first member (n = 1) of the Ruddlesden-Popper $La_{n+1}Ni_nO_{3n+1}$ series and can be described as an alternation along the *c* crystallographic direction of $LaNiO_3$ perovskite blocks and LaO rock-salt type layers [27]. Contrary to most memristive oxides which are primarily oxygen deficient, L2NO4 films exhibit oxygen surplus and are capable of storing a variety of oxygen stoichiometry (δ) [28]. In the La$_2$NiO$_{4+\delta}$ structure, interstitial oxygen serves as a negatively charged dopant, which has to be compensated by the generation of electronic holes to maintain charge neutrality. The presence of highly mobile interstitial oxygen ions in combination with the reactive TiN electrode leads to the creation of a TiN$_x$O$_y$ interlayer at the metal/oxide interface [29]. It was demonstrated that the TiN/L2NO4/Pt devices show non-volatile RS with unique properties, such as a "soft-forming" step [31], which contrary to most filamentary devices does not require the application of a high voltage to initialize the RS process and does not induce binary RS in the device. Thus, the TiN/L2NO4/Pt devices feature gradual analog RS and the ability to update the conductance under the application of the voltage pulses, i.e., long-term potentiation (LTP) and depression (LTD) [29]. At the same time, the use of long voltage pulses (over 50 ms) during the LTP/LTD process shown in previous work leads to a relatively large power consumption, making a reduction in the pulse duration highly desirable. Furthermore, TiN/L2NO4/Pt devices have yet to demonstrate their capability to exhibit STDP.

It was previously shown that the oxygen content in the La$_2$NiO$_{4+\delta}$ film can be modified by post-deposition annealing over a wide range of oxygen stoichiometry, and can influence the RS behavior of planar Ti/L2NO4/Pt devices [28]. By annealing in an inert atmosphere (Ar and H$_2$), it was possible to increase the resistivity of the L2NO4 thin films; on the other hand, the thermal annealing in an oxidizing (O$_2$) atmosphere led to an increase in the conductivity of the L2NO4 film due to the additional generation of the hole carriers. In this work, the influence of thermal annealing on the resistive switching properties of vertical TiN/L2NO4/Pt devices is thoroughly investigated. We show that the annealing in an inert Ar atmosphere increases the initial resistance of the device, resulting in a filamentary behavior with the typical forming step. In addition, the filamentary RS in the Ar annealed sample exhibits LTP/LTD with µs range pulses, allowing to achieve a larger memory window with reduced energy consumption. On the contrary, thermal annealing in an $O_2$ atmosphere leads to forming-free behavior with HRS and LRS resistance states that depend on the device area, which indicates that the interfacial type switching is dominant in the $O_2$ annealed devices. However, the slower switching kinetics of this device leads to an increase in power consumption during the learning process. Nevertheless, both devices demonstrated the ability of LTP/LTD and STDP-based learning, and their potential for neuromorphic applications was evaluated using SNN simulation tools.

## 2. Results & Discussion

### 2.1. Structural characterization of the annealed $La_2NiO_{4+\delta}$ films

**Figure 1**a shows GI-XRD spectra for $La_2NiO_{4+\delta}$ films before and after the annealing in Ar and $O_2$ atmospheres. All films show a crystalline structure where main diffraction peaks are attributed to either the $La_2NiO_{4+\delta}$ tetragonal phase (space group *I4/mmm*; ICDD: 00-034-0314) or the orthorhombic phase (space group *Fmmm*; ICDD: 01-086-8663), as confirmed by the match of the peak positions between L2NO4/Pt samples and those of the database plotted by drop lines in Figure 1a. The Pt diffraction peaks are marked by grey circles. The patterns of both structures are very similar since peak intensities and peak resolution in thin films are too low to distinguish the peak splitting going from tetragonal to orthorhombic phase and, therefore, it is extremely difficult to discriminate between the two phases based on GI-XRD exclusively. Moreover, no decomposition of L2NO4 films is observed after being exposed to the thermal treatment.

To analyze the impact of the annealing process on the crystalline structure, the *a, b* and *c* lattice parameter values were extracted from the position of the GI-XRD peaks in Figure 1a. Despite the possibility of having two phases, for simplicity, only the orthorhombic *Fmmm* structure was chosen to obtain the lattice parameters. The values obtained are: Ar annealed sample ($a = 5.439 \pm 0.005$ Å, $b = 5.492 \pm 0.002$ Å, $c = 12.583 \pm 0.009$ Å), as-deposited sample

($a = 5.439 \pm 0.002$ Å, $b = 5.490 \pm 0.001$ Å, $c = 12.643 \pm 0.008$ Å) and $O_2$ annealed sample ($a = 5.405 \pm 0.007$ Å, $b = 5.467 \pm 0.002$ Å, $c = 12.681 \pm 0.014$ Å). As can be seen in Figure 1b, the value of the lattice parameter $c$ increases when the sample is oxidized, corresponding to the trend previously observed for the epitaxial $La_2NiO_{4+\delta}$/STO structures [28], which indicates the increase of interstitial oxygen concentration after annealing in oxidizing conditions ($O_2$) and the decrease in the inert gas (Ar) atmosphere. In turn, the $a$ and $b$ lattice parameters slightly decrease with the increase in the interstitial oxygen content, similar to those previously reported in the literature [32–34]. These results confirm that the oxygen content is tunable by post-thermal treatments and that it affects the structure and oxidation state of polycrystalline L2NO4 grown on Pt in a similar fashion as for epitaxial L2NO4 [28]. Next, the surface morphology of the annealed films was analyzed using SEM and AFM imaging (Figure S1 in the Supplementary Material). SEM images show that the surface morphology of both films is formed by randomly oriented grains with small grain sizes of the order of 25-45 nm. Both films are homogeneous, polycrystalline, and dense. Moreover, no cracking or pinholes were observed in any of the samples. Both Ar and $O_2$ annealed films showed similar root-mean-square (RMS) values of ~2.9 nm and ~2.8 nm, respectively.

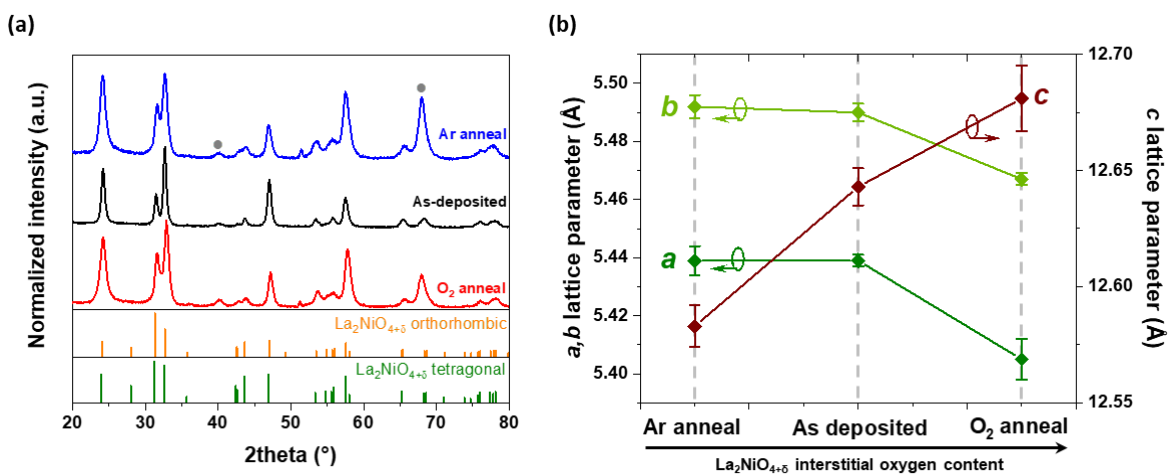

Figure 1. Structural characterization of the annealed L2NO4 films deposited on a platinized silicon substrate. (a) GI-XRD patterns, main diffraction peaks attributed to the either $La_2NiO_{4+\delta}$ tetragonal phase (space group *I4/mmm*; ICDD: 00-034-0314; green lines at the bottom) or the orthorhombic phase (space group *Fmmm*; ICDD: 01-086-8663; orange lines

at the bottom) in polycrystalline form (randomly oriented). The substrate (Pt) peaks are indicated by the grey dots. (b) Evolution of *a*, *b* and *c* lattice parameters with the increase of interstitial oxygen content in the $La_2NiO_{4+\delta}$ films. The lattice parameters were estimated from peak positions in Figure 1a.

Previously it was shown that a $TiN_xO_y$ interlayer forms at the TiN/L2NO4 interface [29], which shows the scavenging ability of the TiN top electrode in contact with the oxygen-rich L2NO4 layer. However, the distribution of oxygen within the L2NO4 layer has not been investigated yet. **Figure 2** shows the analysis of the O-K edge fine structure using monochromated STEM-EELS techniques. For this experiment, two lamellas were prepared from the Ar annealed (Figure 2a) and $O_2$ annealed (Figure 2b) TiN/L2NO4/Pt devices prior to any electrical measurements. The EELS measurements were performed along the direction indicated by arrows in Figure 2a,b (see details in Experimental section). It can be seen from Figures 2c and 2d that the O-K edge spectra look similar in both samples, showing two main features: the main edge at ~531 eV determined by the hybridization of La-O and Ni-O, and the pre-edge peak at ~528 eV (highlighted in grey), associated to hybridized Ni3d-O2p levels [35]. Nakamura et al. have previously shown that the intensity of the pre-edge peak corresponds to the interstitial oxygen concentration in $La_2NiO_{4+\delta}$ [35]. It can be observed that the pre-edge peak intensity is lower in the Ar annealed sample (Figure 2c) compared to the $O_2$ annealed sample (Figure 2d). In addition, the pre-edge peak intensity gradually decreases when moving away from the bottom L2NO4/Pt interface in both samples. Figures 2e and 2f show the comparison between O-K edge EELS spectra measured for the L2NO4/Pt and TiN/L2NO4 interfaces of the Ar annealed and $O_2$ annealed device, respectively, highlighting the difference in the pre-edge peak intensity (region 1 indicated by grey arrow). On the other hand, the intensity of the main edge near the slope at the lower energy side, which is related to the contribution of Ni–O hybridization[36] increases at the TiN/L2NO4 interface compared to the L2NO4/Pt interface (region 2 indicated by green arrow). To estimate qualitatively the differences between the samples, we carried out the analysis of the pre-edge peak intensity following the procedure described by Nakamura et al. [35]: two auxiliary lines, one is the background line and the other is the extension of the slope of the main edge, are drawn near the pre-edge peak. Surrounding area of the pre-edge peak and auxiliary lines, integrated pre-edge peak intensity, is then normalized to the integrated intensity of the whole O-K edge in the range from 527 eV to 555 eV. Figure 2g shows the normalized integrated pre-edge peak intensity of O K-edge as a function of distance from the L2NO4/Pt interface for both Ar and $O_2$ annealed samples. This analysis confirms the existence of a gradient of interstitial oxygen concentration across the whole thickness of the L2NO4 film in both devices, which shows that the scavenging effect from the TiN electrode affects gradually the whole L2NO4 volume and is not localized at the TiN/L2NO4 top interface. Moreover, the offset between the two curves indicates that the concentration of interstitial oxygen ions is higher in the $O_2$ annealed sample across the whole L2NO4 thickness, which is consistent with the shift in the Ni-K edge position previously observed by XANES [28]. Therefore, both effects contribute to the behavior observed in Figure 2g: the scavenging effect from the TiN electrode produces a gradient in oxygen concentration, while the annealing atmosphere affects the interstitial oxygen content ($\delta$) and defines the average stoichiometry in the film.

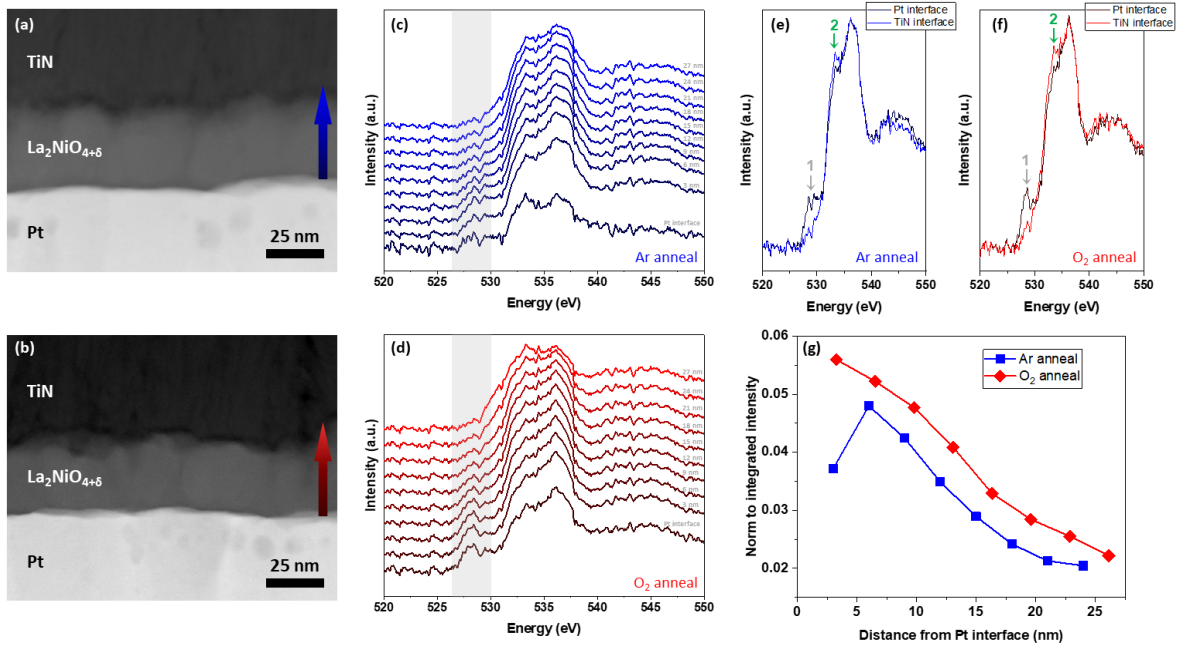

Figure 2. STEM-EELS analysis of O-K edge fine structure in the L2NO4 layer. (a,b) HAADF-STEM cross-sectional image of the (a) Ar annealed and (b) $O_2$ annealed device. The arrows indicate the scanning direction of the EELS profile analysis. (c,d) O-K edge EELS spectra of the (c) Ar annealed and (d) $O_2$ annealed device. Spectra are recorded every 3 nm; numbers indicate the distance to the Pt interface. The pre-edge peak is highlighted. (e,f) O-K edge EELS spectra taken from the L2NO4/Pt and TiN/L2NO4 interfaces of the (e) Ar annealed and (f) $O_2$ annealed device. (g) Normalized integrated pre-edge peak intensity of O K-edge as a function of distance from L2NO4/Pt interface.

## 2.2. Memristive properties of the TiN/L2NO4/Pt devices as a function of the annealing conditions

**Figure 3** shows the contact area dependence of the average initial resistance (IR) of the annealed TiN/L2NO4/Pt devices. For each sample, ten devices per pad size were measured; then average values were calculated for each electrode size and represented by diamonds with error bars. In general, the Ar annealed sample shows higher IR values for all device sizes. The IR differs between the Ar annealed device and the $O_2$ annealed device by approximately one order of magnitude. The higher IR observed in the Ar annealed sample (compared to the oxygen-rich sample) can be attributed to the lower hole carrier concentration and expected higher work function of the L2NO4 film with lower oxygen content. On the other hand, the increase in the interstitial oxygen concentration in $O_2$ annealed L2NO4 films generates more holes in the bulk, leading to lower resistivity in the oxidized sample [28]. This result shows that the device's IR can increase when subjected to a reducing atmosphere, resulting in a higher resistance, which could help to reduce the leakage current of TiN/L2NO4/Pt devices. It can be also seen from Figure 3 that the devices with the contact area of 50x50 µm² show the lowest variability of IR; therefore, they were chosen for further investigation of the RS behavior of the annealed devices.

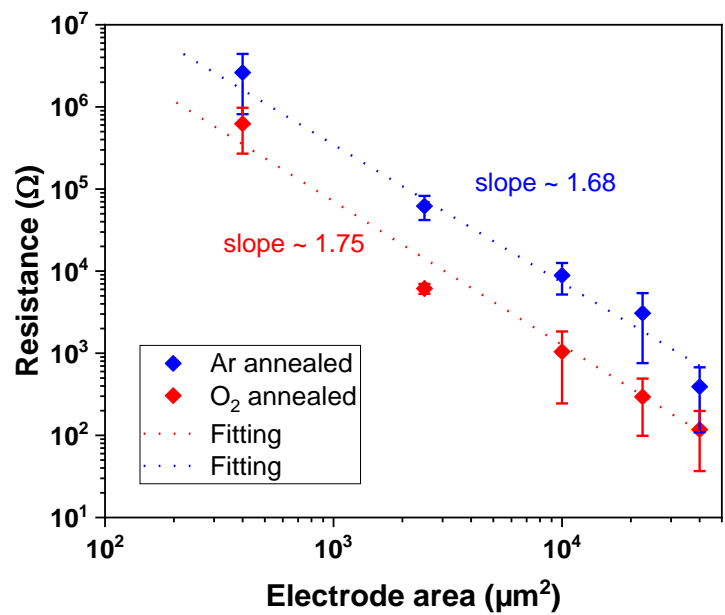

Figure 3. Initial resistance values of TiN/L2NO4/Pt devices as a function of the electrode area for Ar and $O_2$ annealed samples. For each sample, 10 different electrodes have been measured per pad size. $V_{READ} = 0.1$ V The dotted lines were obtained by linear fitting of the experimental data.

**Figure 4** shows the *I-V* characteristic averaged over 60 consecutive cycles measured for the annealed TiN/L2NO4/Pt devices. Both Ar annealed (Figure 4a) and $O_2$ annealed (Figure 4b) devices exhibit similar counter-eightwise RS behavior with the SET process at the negative polarity and the RESET process at the positive polarity, which had been previously observed for as-deposited TiN/L2NO4/Pt devices [29]. The Ar annealed device exhibits an abrupt resistance decrease during the first application of the negative bias of $V_F \approx -2.5$ V, which resembles the "soft-forming" step observed in the as-deposited TiN/L2NO4/Pt devices [31] required to induce the RS behavior (Figure S2a in the Supplementary material). Figure 4c shows that the HRS and LRS resistance do not depend on the device's area, which suggests a filamentary-like RS since in that case the resistance should be determined by the size of the conductive filament. However, despite the presence of the initial "soft-forming" step, the Ar annealed sample shows gradual RS, so the device can potentially exhibit analog properties. On the other hand, from Figure S2b in the Supplementary material it can be seen that the first RS cycle of the O$_2$ annealed device shows a change in current similar to the "soft-forming" step when the voltage in negative polarity reaches $V_F \approx$ -2.5 V. However, in this case the HRS and LRS resistance depend on the area of the device (Figure 4d), suggesting interfacial-like RS behavior. In turn, for larger devices (device area over 85x85 µm$^2$), the RS window almost disappears most likely because of the strong increase in current.

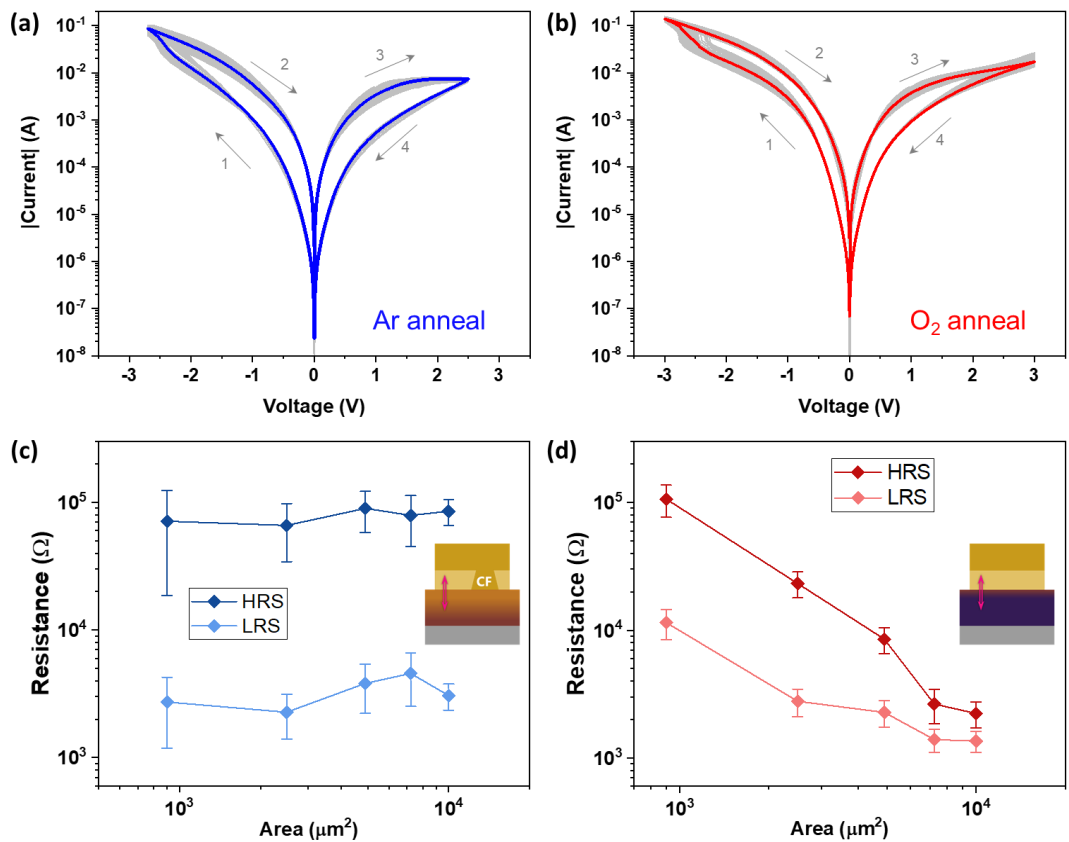

Figure 4. I-V characteristics of (a) the Ar annealed and (b) the O$_2$ annealed TiN/L2NO4/Pt devices averaged over 60 consecutive cycles (grey lines: second to 60th DC sweep; mean lines are shown in blue and red, respectively). The dependence of the HRS and LRS resistance on the area of the device of (c) the Ar annealed and (d) the O$_2$ annealed TiN/L2NO4/Pt devices. Each point is averaged over 5 devices. $V_{READ}$ = 0.1 V. Inset shows the device stacks from top to bottom: Pt/L2NO4/TiN$_x$O$_y$/TiN with the sketch of the proposed switching mechanisms: filamentary for Ar annealed device and interfacial for O$_2$ annealed device. The arrow describes oxygen exchange between L2NO4 and TiN$_x$O$_y$ layers under electric bias.

To compare the cycle-to-cycle (c2c) variability of the devices, we performed cumulative probability (CP) calculations. **Figure 5**a and 5b show the CP plots based on the 60 consecutive cycles measured from one device for Ar annealed and O$_2$ annealed devices, respectively. It can be seen that both devices show good c2c variability in HRS, with σ = 12.8% and 11% for Ar and O$_2$ annealed devices, respectively. On the other hand, in the LRS, the Ar annealed device shows a low variability of σ = 6.7 %, while the O$_2$ annealed device exhibits higher c2c variability of σ = 55%. Next, to confirm the observed behavior of the annealed TiN/L2NO4/Pt devices and assess device-to-device (d2d) variability, 40 devices (50x50 µm$^2$) were measured for each sample. Figure 5c shows the box plot of the resistance values in HRS and LRS for Ar and O$_2$ annealed devices. Simultaneously, Figure 5d shows the SET and RESET voltage distributions as well as the forming voltage distribution for the Ar annealed device during the RS process. It can be seen that the d2d variability of resistance and voltage is similar for the two types of annealing, although the LRS variability of O$_2$ annealed device is slightly higher. Interestingly, this behavior is atypical for RS devices, since the filamentary devices tend to show more variability [37]. The mean LRS and HRS resistance values are 2 kΩ and 58 kΩ for the Ar annealed device, respectively, resulting in the mean memory window of ~28 with the mean $V_{SET}$ = -2.7 V and $V_{RESET}$ = 2.5 V. The mean forming voltage value $V_F$ = -2.7 V is equal to the mean $V_{SET}$ for the Ar annealed device, however, the variability of the $V_F$ is slightly higher which is usually related to the stochastic nature of the CF formation process [38]. For the O$_2$ annealed device, the mean LRS and HRS resistance values are 2.4 kΩ and 31 kΩ, respectively, resulting in the memory window of ~13 with the mean $V_{SET}$ = -3 V and $V_{RESET}$ = 3 V. Therefore, overall, the Ar annealed device demonstrated a larger memory window with the use of a smaller voltage amplitude during standard RS.

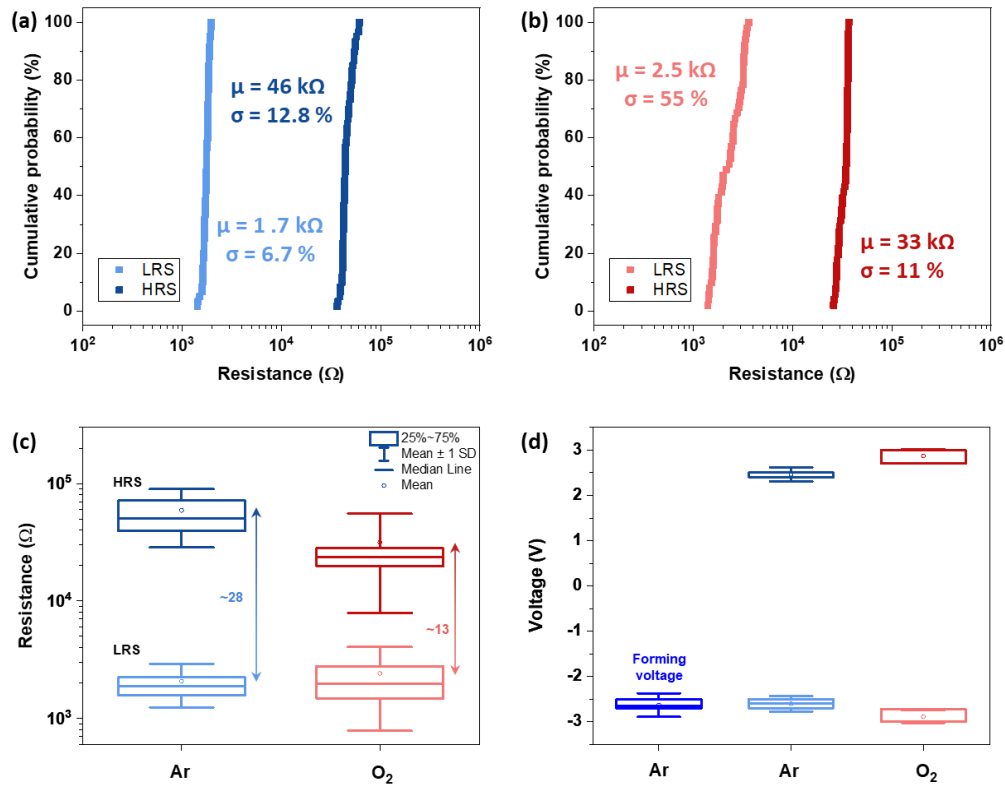

Figure 5. Cycle-to-cycle variability of HRS and LRS resistance for (a) the Ar annealed and (b) the O$_2$ annealed TiN/L2NO4/Pt devices based on 60 DC sweeps. $V_{READ}$ = 0.1 V. Box plot of (c) HRS and LRS resistance and (d) SET and RESET voltage (+forming voltage for Ar annealed sample) for annealed TiN/L2NO4/Pt devices based on the mean values from 40 separate devices (50x50 µm$^2$).

*2.3 Synaptic properties of the TiN/L2NO4/Pt devices as a function of the annealing conditions*

To evaluate the influence of the oxygen stoichiometry of the L2NO4 films on the analog properties of the TiN/L2NO4/Pt devices, LTP/LTD were tested in both Ar and O$_2$ annealed devices by the application of trains of voltage pulses with fixed amplitude and duration. **Figure 6**a shows 10 subsequent LTP/LTD curves for the Ar annealed TiN/La$_2$NiO$_{4+\delta}$/Pt device. For potentiation (depression), 100 pulses with a voltage amplitude $V_{SET}$ = -2.7 V ($V_{RESET}$ = +2.3 V) and a duration of 10 µs (1 µs) were applied. The voltage amplitude and duration were optimized beforehand so that a stable memory window could be achieved over subsequent LTP/LTD cycles with minimal power consumption. It can be observed that the conductance of the Ar annealed device changes gradually with the application of the identical pulses between $G_{min} \approx 2 \times 10^{-5}$ S and $G_{max} \approx 4 \times 10^{-5}$ S, resulting in a memory window of ~2. Moreover, the initial $G_{min}$ and $G_{max}$ values can be maintained throughout the potentiation/depression cycling process within a reasonable margin, although the memory window slightly decreases to ~1.83 at the last cycle. Next, the same testing procedure was applied to the $O_2$ annealed device (Figure 6b). For potentiation (depression), 100 pulses with a voltage amplitude $V_{SET} = -2.7$ V ($V_{RESET} = +2.7$ V) and a duration of 1 ms (1 ms) were used. Similar to the Ar annealed device, the gradual change of conductance between $G_{min} \approx 4.3 \times 10^{-5}$ S and $G_{max} \approx 5.3 \times 10^{-5}$ S, however, the resulting memory window is narrower (~1.23). It can be observed that even though longer pulses were applied to the $O_2$ annealed device, the memory window is smaller compared to the Ar annealed device. The observed difference can be explained by the presence of Joule heating associated with the CF in the case of the Ar annealed device, which is crucial for the improvement of the switching kinetics [39,40]. The current-induced local increase of the temperature in the CF region accelerates the drift and diffusion of oxygen ions, therefore, less energy is needed to cause a conductance change in the device [41]. At the same time, we can conclude that both devices can demonstrate LTP/LTD and therefore are suitable for use as artificial synapses in an SNN architecture.

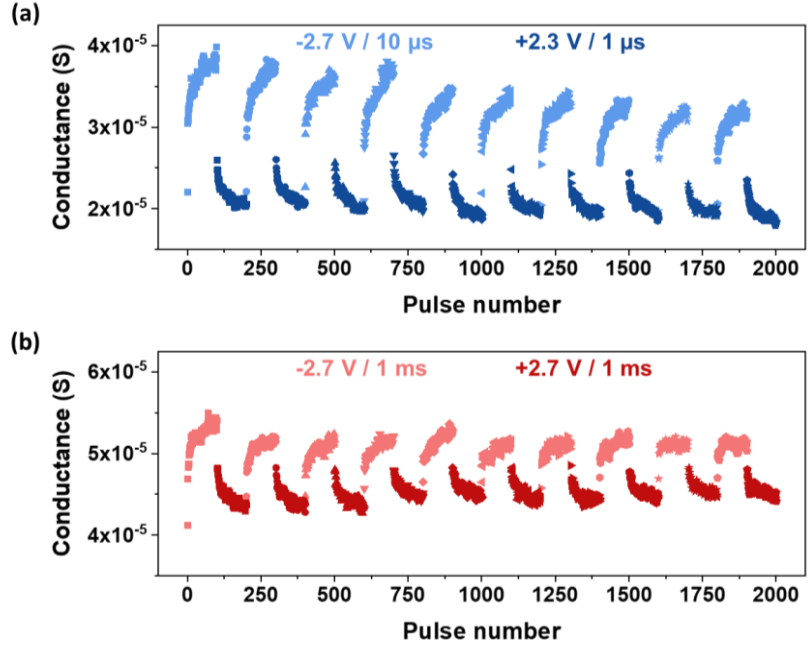

Figure 6. Reversible LTP/LTD for the (a) Ar annealed and (b) $O_2$ annealed TiN/L2NO4/Pt devices. The characteristics of voltage pulses are indicated in each figure. The amplitude/duration of the read voltage pulses is 0.1 V/10 ms in both cases.

The use of the shorter pulses during the LTP/LTD process allows for a decrease in the energy consumption in the learning process. The energy consumed per spike for LTP and LTD in Figure 6 was estimated as

$$E_{LTP} = V_{SET}^2 \times G_{min} \times \Delta t_{LTP} \quad (1)$$

$$E_{LTD} = V_{RESET}^2 \times G_{max} \times \Delta t_{LTD} \quad (2)$$

where *V* is the writing voltage amplitude and *Δt* is the pulse duration used during potentiation or depression transition [42]. The results of the calculations for Ar and $O_2$ annealed samples are summarized in **Table 1**. In addition, the energy value achieved with the TiN/L2NO4/Pt device based on the as-deposited L2NO4 film as well as LTD/LTP pulse parameters from [29] are included in Table 1 for comparison. It can be seen that the use of the pulses in the µs range in the case of the Ar annealed sample allowed to decrease energy consumption down to ~1.5 nJ/spike for LTP and ~0.2 nJ/spike for LTD. While these values still do not reach the fJ-pJ range achieved in other neuromorphic synapses [43–45], it is a drastic improvement compared to the energy consumption of the as-deposited TiN/L2NO4/Pt device, allowing a reduction of 3 and 5 orders of magnitude for LTP and LTD, respectively. In the case of the $O_2$ annealed device, the energy consumption is ~292 nJ/spike for LTP and ~401 nJ/spike for LTD. Further reduction in energy consumption for the $O_2$ annealed sample can be achieved by the decrease in the device area since it brings an HRS and LRS resistance increase.

To evaluate the potential of the devices under study as the artificial synapse in SNN architectures and estimate the influence of the observed changes in the LTP/LTD behavior (Figure 6) on the learning accuracy, we used an in-house SNN simulation tool [46]. For this, the experimental data in Figure 6 were fitted using an asymmetric conductance model [47], where different fitting functions are used for the conductance change during potentiation (3) and depression (4):

$$\left.\frac{dG}{dt}\right)_+ = f_+(V)e^{-\beta_+\frac{G-G_{min}}{G_{max}-G_{min}}} \qquad (3)$$

$$\left.\frac{dG}{dt}\right)_- = -f_-(V)e^{-\beta_-\frac{G_{max}-G}{G_{max}-G_{min}}} \qquad (4)$$

where $\beta_+$ and $\beta_-$ are fitting parameters. Next, the extracted behavior is implemented as a synaptic plasticity in the SNN tool which is later trained on the MNIST dataset consisting of black-and-white images of handwritten digits with a resolution of 28x28 pixels. The training process algorithm is based on the work of Diehl and Cook [4] with the base accuracy for the SNN under study being 86% at the algorithmic level. A detailed explanation of the training process with the SNN tool can be found elsewhere [29,46]. Table 1 shows the learning accuracy obtained by implementing the first cycle of the potentiation/depression cycling process for Ar and $O_2$ annealed devices as a synaptic weight update. In addition, the accuracy value achieved with the TiN/L2NO4/Pt device based on the as-deposited L2NO4 film adapted from [29] is included in Table 1 for comparison. It can be seen that the use of shorter voltage pulses in the case of the Ar annealed device allowed it to reach 81.9% accuracy, which is a large improvement compared to the as-deposited TiN/L2NO4/Pt device where the longest pulses were used. For the $O_2$ annealed device, the learning accuracy value is 79.3%, which shows that the learning accuracy in our case depends on the pulse length. Therefore, minimizing energy consumption also leads to the improvement in the learning accuracy of the SNN.

Table 1: Comparison of the energy consumption during the LTP/LTD process and learning accuracy estimated for an L2NO4-based SNN trained using STDP for MNIST database (base accuracy of the tool under ideal conditions 86%)

| Device | LTP/LTD pulse parameters | Energy consumption for LTP/LTD, nJ/spike | Accuracy | Ref. |
|---|---|---|---|---|
| Ar annealed | -2.7 V 10 µs / +2.3 V 1 µs | 1.5 / 0.2 | 81.9 % | This work |
| $O_2$ annealed | -2.7 V 1 ms / +2.7 V 1 ms | 292/ 401 | 79.3% | This work |
| As-deposited | -1.25 V 250 ms / +2.0 V 250 ms | 5580 / 25000 | 76.8 % | [29] |

Another way of emulating synaptic functions is to show the STDP behavior directly in the memristive devices. In this rule, the synaptic weight update depends on the relative timing between post- and pre-synaptic spikes. For this test, we used the bio-inspired waveform for spike emulation consisting of two triangular pulses of opposite polarity (**Figure 7**). The spike is formed from three components: (1) positive ramp from 0 V to $V_1 = 2$ V ($t = 100$ µs); (2) negative ramp from $V_1 = 2$ V to $V_2 = -0.7$ V ($t = 100$ µs); (3) positive ramp from $V_2 = -0.7$ V to 0 V ($8t = 800$ µs). The first triangle emulates the action potential of a neuron and the second triangle emulates a refractory period when a neuron returns to the resting state [48]. In the case of potentiation, the pre-synaptic spike arrives before the post-synaptic spike, so the time difference $\Delta t$ is positive ($\Delta t > 0$). In the case of depression, the post-synaptic spike arrives before the pre-synaptic spike, so the time difference $\Delta t$ is negative ($\Delta t < 0$). Consequently, the size of the potentiation (depression) area above $V_{th} = 2$ V is determined by the $\Delta t$. The schematics with examples of residual pulses for potentiation and depression are shown in Figure 7. To sum up, the waveform used in this study is characterized by 3 main parameters: $V_1$, $V_2$ and $t$.

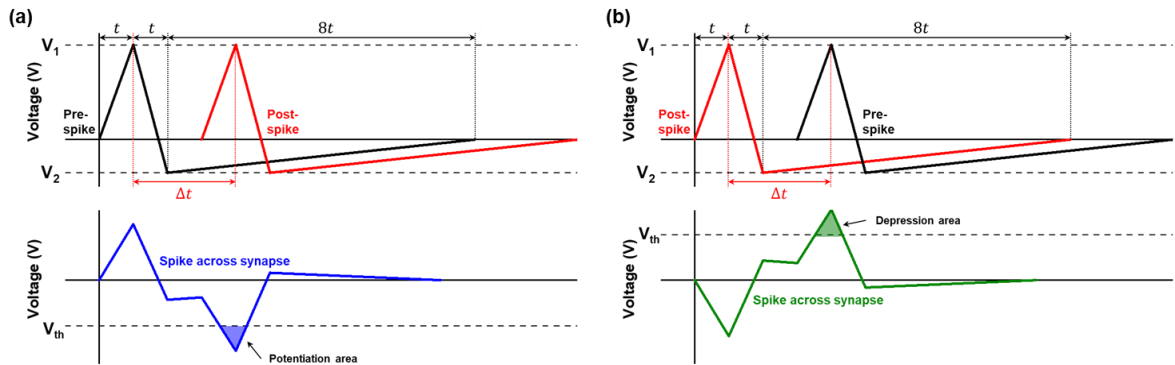

Figure 7. The waveforms used for the STDP measurements. The bottom graphs show examples of the residual waveforms which are obtained by the subtraction of the post-synaptic pulse from the pre-synaptic. (a) In the case of potentiation, the pre-synaptic spike arrives before the post-synaptic spike, so the time difference $\Delta t$ is positive ($\Delta t > 0$). (b) In the case of depression, the post-synaptic spike arrives before the pre-synaptic spike, so the time difference $\Delta t$ is negative ($\Delta t < 0$).

**Figure 8** shows the relationship between relative conductance change $dG/G_0$ and the time difference between post- and pre-synaptic spikes $\Delta t$. The data in Figure 8 is averaged over 3 devices for both samples. The experimental results can be fitted by the Hebbian asymmetric learning rule:

$$\frac{dG}{G_0} = \alpha \times e^{-\frac{\Delta t}{\tau}} \quad (5)$$

where $\alpha$ and $\tau$ are the scaling factor and the time constant, respectively. It can be observed that the Hebbian asymmetric learning rule can be implemented in both devices. For the Ar annealed device (Figure 8a), which exhibits filamentary-like switching, an asymmetric behavior can be observed, where the potentiation is stronger than depression. Similar behavior is reported for other types of filamentary-RS switching oxides such as $WO_x$ [49] and $HfO_2$ [50]. On the other hand, the O$_2$ annealed device (Figure 8b) shows a more symmetric response, however, the relative change in conductance $dG/G_0$ is much smaller for both potentiation and depression compared to the Ar annealed device. Since the LTP/LTD test showed that the $O_2$ annealed device requires significantly more energy for the weight update (Table 1), at the same pulse parameters the weight update is expected to be smaller. Overall, this shows that the STDP behavior can be observed in TiN/L2NO4/Pt devices in both filamentary and interfacial switching modes. In addition, we also investigated the influence of the voltage amplitude and time on the shape of the STDP curves (details of this experiment can be found in section 3 in the Supplementary material).

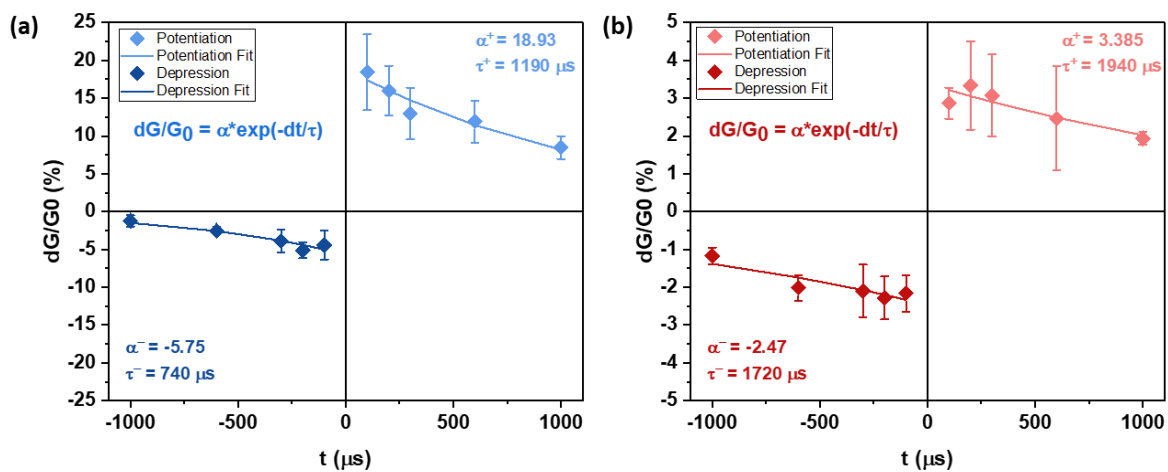

Figure 8. Experimental demonstration of spike-timing-dependent plasticity (STDP) of (a) Ar annealed and (b) $O_2$ annealed TiN/L2NO4/Pt devices. It should be noted that the $dG/G_0$ axis scale is different for each plot.

The fitting parameters obtained for post-synaptic and pre-synaptic weight updates enable the direct use of the STDP curves extracted from the memristive devices within SNN simulation tools. To analyze the potential behavior of TiN/L2NO4/Pt synapses in the SNN, we implemented our fitted learning rules in the simplified version of the unsupervised learning architecture described in Diehl and Cook [4] from the BindsNET package [51].

**Figure 9**a shows the schematic illustration of the SNN simulation tool workflow. The SNN architecture consists of two layers: the input layer and the processing layer. Similar to the in-house SNN simulation tool described above, here we also based our training on 60000 images from the MNIST database. Each pixel is represented by one input neuron; therefore,

the first layer consists of 28x28 = 784 input neurons. The processing layer, in turn, consists of the excitatory and inhibitory neurons which are connected in a one-to-one fashion to induce the lateral inhibition of the learning process and enable the winner-take-all (WTA) mechanism. All the inhibit neurons are fully reconnected to the excitatory neurons except for the self-inhibition path. This way, once one output neuron fires, it suppresses the membrane potential of the other excitatory neurons. In our approach, the synaptic weight update is based on the STDP rule which is extracted from the fitting of experimental data in Figure 8 with the addition of the given minimum and maximum conductance of the synapse. Therefore, the pre-synaptic and the post-synaptic conductance change in a synapse over time (t) can be expressed as follows:

$$\Delta G_{post} = G_{max} - \alpha_{post} \times e^{\frac{t}{\tau_{post}}} \qquad (6)$$

$$\Delta G_{pre} = G_{min} + \alpha_{pre} \times e^{-\frac{t}{\tau_{pre}}} \qquad (7)$$

where $G_{min}$ and $G_{max}$ are the minimum and maximum synaptic conductance and their values are extracted from LTP/LTD curves in Figure 6. During the online unsupervised learning process, each excitatory neuron learns one digit from 0 to 9 through the WTA mechanism. After training is done, a class is assigned to each neuron, based on its highest response to the ten classes of digits over one presentation of the training set [4]. An example of the resulting rearranged input to excitatory neuron weights in the case of 100 neurons in the processing layer as well as the grid of corresponding categorical assignments is also shown in Figure 9. Next, in inference testing performed on a separate dataset of 10000 digits, the network sums up the spike count of excitatory neurons for each class label and classifies an input based on the label that accumulates the most spikes. The inference accuracy is then calculated based on the percentage of successful digit recognition. The number of excitatory and inhibitory neurons can be varied and was previously shown to impact the network's accuracy [52]. Figure

9b shows that the inference accuracy in our case also depends on the number of exhibitory/inhibitory neurons; at the same time, the two devices under study (Ar and $O_2$ annealed) provide similar accuracy when used as synapses in an STDP-trained SNN. In addition, we also show that the accuracy can be maintained with STDP curves of different time and voltage resolutions as long as the shape of the curve follows the exponential rule (details of this experiment can be found in section 3 in the Supplementary material).

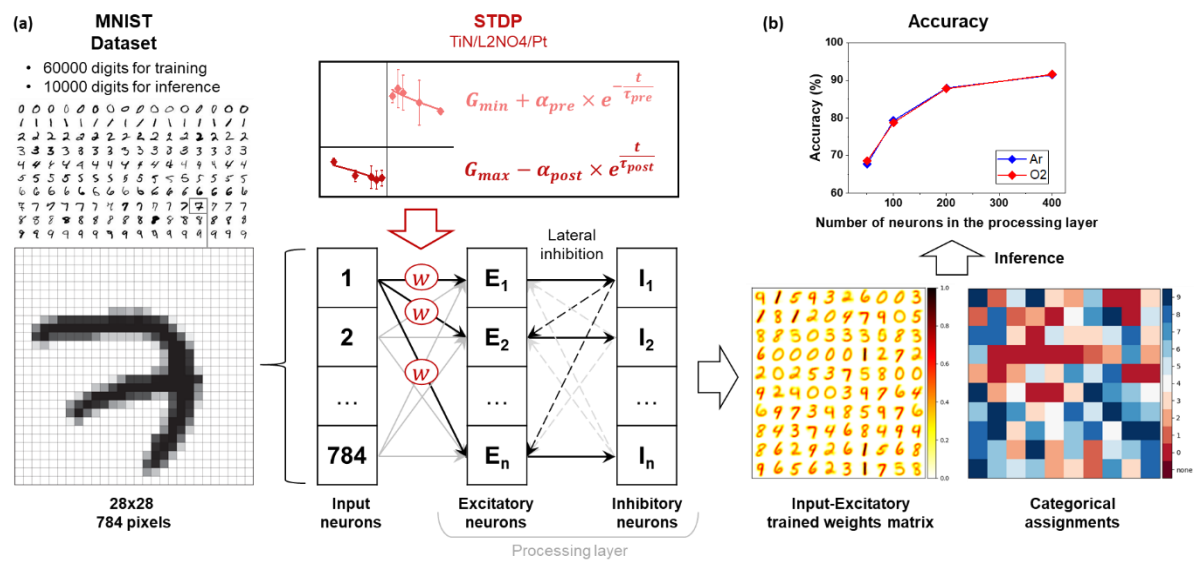

Figure 9. (a) The schematic illustration of the Bindsnet SNN architecture. The STDP equations obtained by fitting the experimental data in Figure 8 are used as weight update functions for synaptic connections between input and excitatory neurons. (b) The inference accuracy of the SNN based on the experimental STDP rules of Ar and $O_2$ annealed TiN/L2NO4/Pt devices as a function of the number of neurons.

## 2.4 Challenges and perspectives of the TiN/La$_2$NiO$_{4+\delta}$/Pt devices

Our study shows that the control of the oxygen content in the $La_2NiO_{4+\delta}$ film allows for the fine tuning of the memristive and synaptic behavior of the TiN/La$_2$NiO$_{4+\delta}$/Pt devices. However, even though the simulation results show that these devices are promising for artificial synapse applications, the potential technological integration of these devices faces a number of challenges. First, both deposition and annealing temperatures used in this study exceed the maximum temperature of 450 °C defined for CMOS back-end-of-line (BEOL) integration.

While the post-deposition annealing temperature of 500 °C could be easily reduced to 450 °C, obtaining the same composition of $La_2NiO_{4+\delta}$ films at lower deposition temperatures has proven to be challenging. Our previous study showed that at the MOCVD temperature of 500 °C the films are amorphous [26]. The crystallization of the amorphous films at BEOL-compatible temperatures to the $La_2NiO_{4+\delta}$ phase is difficult since this phase is known to stabilize at high temperatures [53]. One of the possible solutions can be laser annealing, which allows for local heating of the film and therefore can be used in BEOL technology [54]. Alternatively, one could approach different deposition methods. For example, Sønsteby et al. used atomic layer deposition (ALD) to obtain $LaNiO_3$ epitaxial films on $SrTiO_3$ substrates at T = 225 °C with no annealing required [55]. In the future, it would be interesting to try a similar approach to achieve the $La_2NiO_{4+\delta}$ phase at BEOL-compatible temperatures. Another challenge is the scalability of the $TiN/La_2NiO_{4+\delta}/Pt$ devices, which is especially important for high-density artificial synapse applications. The volume of solely the switching oxide of the studied devices is 30 nm x 50 µm x 50 µm, which is very large compared to device sizes of nm-scale already combined in CMOS-based circuits by co-integration [56–59]. Therefore, the miniaturization of these devices represents a significant developmental objective to be prioritized moving forward. Integration of TiN/L2NO4/Pt devices into cross-point and cross-bar architectures will facilitate further downscaling, advancing these devices closer to technological integration. Lastly, while the simulation tools provide valuable information about the potential behavior of the devices, the direct measurements on hardware are more reliable. The cross-bar architecture integration will also allow for the investigation of a small-scale L2NO4-based hardware SNN and further verification of the device behavior as an artificial synapse. Moreover, the shift from the individual devices to the cross-bar arrays would speed up the electrical characterization, enabling the collection of more significant statistical data.

## 3. Conclusions

We investigated how the memristive and synaptic properties of TiN/La$_2$NiO$_{4+\delta}$/Pt depend on the oxygen content in the La$_2$NiO$_{4+\delta}$ film. GI-XRD and EELS analysis show that the concentration of the interstitial oxygen ions can be tuned using thermal annealing in the inert gas (Ar) or oxidizing atmosphere (O$_2$). In addition, in both samples, the scavenging effect from the TiN electrode creates a gradient in the interstitial oxygen content within the La$_2$NiO$_{4+\delta}$ layer. Both Ar and O$_2$ annealed devices demonstrate counter-eightwise RS; they also show the ability to achieve reproducible LTP/LTD and STDP behavior based on the Hebbian asymmetric learning rule. At the same time, the Ar annealed device showed filamentary-like RS with HRS and LRS independent of the device area. Moreover, with the Ar annealed device a larger memory window in the LTP/LTD process was achieved at the reduced energy consumption. On the other hand, the O$_2$ annealed device presented slower switching kinetics, which can be attributed to the interfacial-like RS with the area dependence of HRS and LRS resistance. Lastly, the potential behavior of TiN/La$_2$NiO$_{4+\delta}$/Pt synapses in the SNN architecture was studied using simulation tools. For this test, we implemented the learning rules directly measured from the devices as weight update functions between input and processing layer neurons. We showed that in the SNN based on the TiN/La$_2$NiO$_{4+\delta}$/Pt synapses reduction of the energy consumption correlates with improved learning accuracy.

## 4. Experimental section

*Device fabrication:* The L2NO4 films were grown by PI-MOCVD at 600 °C on 1x1 cm$^2$ substrates which consist of Pt (100 nm)/TiO$_2$ (20 nm)/SiO$_2$ (500 nm)/Si (750 µm)/SiO$_2$ (500 nm) structures using La(tmhd)$_3$ (tris(2,2,6,6-tetramethyl-3,5-heptanedionato) lanthanum (III)) and [Ni(tmhd)$_2$] (bis(2,2,6,6-tetramethyl-3,5-heptanedionato) nickel (II)) precursors mixed with m-xylene (1,3-dimethylbenzene) in a 0.02 M solution. The La$_2$NiO$_{4+\delta}$ growth conditions used in this work are detailed in our previous works[29,31] and correspond to the optimized deposition conditions for $La_2NiO_{4+\delta}$ thin films grown on Pt [26]. After the deposition, the films were subjected to thermal annealing at 500 °C for 1 hour in an $O_2$ or Ar atmosphere for oxidizing and reducing of L2NO4 films, respectively. The thickness of the films (~33 nm) was estimated using cross-sectional transmission electron microscopy (TEM).

The microfabrication of the devices was carried out in the PTA clean-room facilities (Grenoble), including laser lithography (Heidelberg instrument µPG 101) and TiN deposition by reactive sputtering equipment (PVD 100 Alliance Concept, the deposition rate is 0.2 nm/s). The top 100 nm-thick TiN square contacts with dimensions of 20 – 200 µm were formed through the lift-off process.

*Structural characterization:* For the phase identification of the L2NO4 thin films, X-ray diffraction in grazing incident mode (GI-XRD) was carried out in a 5-circle Rigaku Smartlab diffractometer. To determine the cell parameters of L2NO4, the XRD patterns were refined using the TOPAS software. The surface morphology was studied by scanning electrode microscopy (SEM) in a FEG ZEISS GeminiSEM 300 microscope. The surface roughness was investigated by atomic force microscope (AFM) in an AFM D3100 Veeco Instrument in tapping mode with a $Si_3N_4$ tip probe. The TiN/$La_2NiO_{4+\delta}$/Pt devices were characterized locally by scanning transmission electron microscopy (STEM) in a probe-corrected FEI Titan 60-300. This instrument is equipped with a CEOS aberration corrector for the condenser system, providing a probe size of ~1 Å, a high-brightness Schottky field emission gun (X-FEG), a Wien filter monochromator and a GIF Tridiem 866 ERS from Gatan for electron energy loss spectroscopy (EELS). High-angle annular dark field (HAADF) images were collected in STEM. Monochromated EELS was performed to obtain high-energy resolution EELS spectra of the O-K edges. The full width at half maximum (FWHM) of the zero-loss peak was ~0.3 eV and the energy dispersion of the spectrometer was 0.1 eV. STEM-EELS spectrum images were collected, 40 x 40 pixels in size with a pixel size of 1.5 nm. EELS spectrum line profiles across the interfaces were obtained with high statistics by integrating 40 spectra (parallel to the interfaces) and binning x2 perpendicular to the interfaces, to obtain a final spatial resolution of 3 nm. TEM lamella specimens were prepared in FEI Helios 650 Dual Beam equipment.

*Electrical characterization:* The complete TiN/$La_2NiO_{4+\delta}$/Pt devices were electrically characterized using a Keithley 4200 semiconductor parameter analyzer. In all measurements described in this work, top electrodes with dimensions of 50×50 µm$^2$ were used (if not specified otherwise). The voltage was applied to TiN electrode while Pt electrode was grounded. The current compliance of $I_{CC} = 0.1$ A was used during device initialization for both polarities to prevent irreversible dielectric breakdown. In the pulse mode, the parameters used for the write and read pulses are varied depending on the tests, which is detailed individually in the description of each test.

**Acknowledgments**

This project has received financial support from the CNRS through the MITI interdisciplinary programs and was partially supported by ANR France within the EMINENT Project ANR-19-CE24-0001 and by the LabEx Minos ANR- 10-LABX-55-01 (for T.-K.K.) and from the grants PID2020-112914RB-I00 funded by MCIN/AEI/10.13039/501100011033, and E13_23R funded by the Aragon regional government. This research has benefited from characterization equipment of the Grenoble INP–CMTC platform supported by the Centre of Excellence of Multifunctional Architectured Materials "CEMAM" n°ANR-10-LABX-44-01 funded by the "Investments for the Future" Program. Part of this work, carried out on the Platform for Nanocharacterization (PFNC), was supported by the "Recherches Technologiques de Base" program of the French National Research Agency (ANR). In addition, this work has been performed with the help of the "Plateforme Technologique Amont (PTA)" in Grenoble, with the financial support of the "Nanosciences aux limites de la Nanoélectronique" Foundation and CNRS Renatech network. The authors acknowledge the use of instrumentation as well as the technical advice provided by the National Facility ELECMI ICTS, node «Laboratorio de Microscopias Avanzadas (LMA)» at «Universidad de Zaragoza». The authors acknowledge Valentine Bolcato (PTA, Grenoble INP) for the deposition of the TiN layer and Sergio Vinagrero Gutierrez (Grenoble INP, TIMA) for assistance with the simulation platform.

**Table of contents**

This article presents a study on the optimization of the TiN/La$_2$NiO$_{4+\delta}$/Pt artificial synapses through control of the La$_2$NiO$_{4+\delta}$ oxygen content by annealing in inert (Ar) or oxidizing (O$_2$) atmosphere. This method allows for fine tuning of the memristive and synaptic properties of the device. Simulations indicate potential of these devices for energy-efficient learning in spiking neural networks.

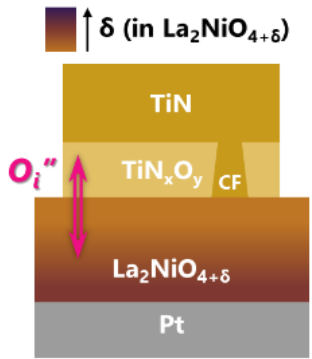 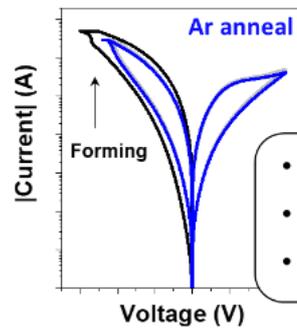

- Fast switching
- Energy efficient
- Filamentary RS

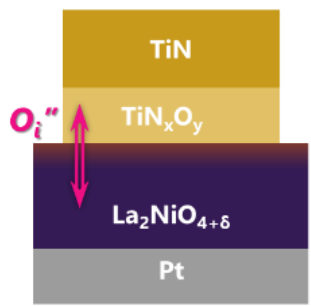 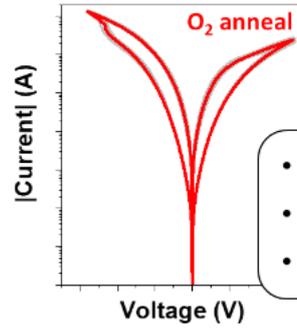

- Forming-free
- Slow kinetics
- Interfacial RS